\documentclass[10pt,twocolumn,twoside]{IEEEtran}
\usepackage[utf8]{inputenc}
\usepackage{color}
\usepackage{multirow}
\usepackage{hyperref}
\usepackage{graphicx}
\usepackage{lmodern}
\usepackage{geometry}
\usepackage{dingbat}
\usepackage{pifont}
\usepackage{balance}
\usepackage{textcomp}
\usepackage{siunitx}
\usepackage{hyperref}
\usepackage{url}
\usepackage{amsmath}
\usepackage{rotating}
\usepackage{booktabs}
\usepackage[noabbrev]{cleveref}
\usepackage{enumitem}
\usepackage{authblk}

\newcommand{\cmark}{\ding{51}}
\newcommand{\xmark}{\ding{55}}

\title{Inline Detection of DGA Domains\\Using Side Information}

\author[1]{Raaghavi Sivaguru}
\author[2,3]{Jonathan Peck}
\author[4]{Femi Olumofin}
\author[1]{Anderson Nascimento}
\author[1,2]{Martine De Cock}

\affil[1]{School of Engineering and Technology, University of Washington, Tacoma, USA}
\affil[2]{Department of Applied Mathematics, Computer Science and Statistics, Ghent University, Ghent, 9000, Belgium}
\affil[3]{Data Mining and Modeling for Biomedicine, VIB Inflammation Research Center, Ghent, 9052, Belgium}
\affil[4]{Infoblox, Santa Clara, USA}

\date{}

\begin{document}

\maketitle

\begin{abstract}
Malware applications typically use a command and control (C\&C) server to manage bots to perform malicious activities. Domain Generation Algorithms (DGAs) are popular methods for generating pseudo-random domain names that can be used to establish a communication between an infected bot and the C\&C server. 
In recent years, machine learning based systems have been widely used to detect DGAs. There are several well known state-of-the-art classifiers in the literature that can detect DGA domain names in real-time applications with high predictive performance. However, these DGA classifiers are highly vulnerable to adversarial attacks in which adversaries purposely craft domain names to evade DGA detection classifiers.

In our work, we focus on hardening DGA classifiers against adversarial attacks. To this end, we train and evaluate state-of-the-art deep learning and random forest (RF) classifiers for DGA detection using side information that is harder for adversaries to manipulate than the domain name itself. Additionally, the side information features are selected such that they are easily obtainable in practice to perform inline DGA detection. The performance and robustness of these models is assessed by exposing them to one day of real-traffic data as well as domains generated by adversarial attack algorithms. We found that the DGA classifiers that rely on both the domain name and side information have high performance and are more robust against adversaries.


\end{abstract}

%
%

\section{Introduction}

Domain Generation Algorithms (DGAs) are subroutines that generate pseudo-random combinations of characters or words, and output domain name strings \cite{plohmann2016comprehensive}. DGAs often use a seed input such as a number, which is embedded as part of the code, or a time-based element such as the system date, time etc., or a combination of both, to generate random strings. These strings are then concatenated with an available top level domain (TLD) to form domain names. The key idea behind DGAs is to generate the same set of domain names when executed by two different machines, such as by a botmaster and on an infected machine, at a given time. The botmaster registers one of the generated domain names, while the infected machines systematically query the domains from the generated list until one of them is resolved. The domains from the list that have not been registered by the botmaster will typically result in an NXDomain (non-existent domain) response when queried, and can be discarded by the infected machine. This technique is often used by a command and control (C\&C) center and an infected bot to establish communication and perform malicious activities as instructed by the C\&C server.


Once communication between the infected machines and the botmaster has been established, the C\&C server can issue commands to the bots to perform malicious activities such as distributed denial of service (DDoS) attacks, spamming, stealing sensitive information from the compromised machines, etc. In the past, malware authors used a predefined list of domain names, which was embedded in the malware, to communicate with the bots. This technique made it easy for the defenders to blacklist the malicious domain names and block further communication, effectively rendering the malware useless. To overcome this, modern C\&Cs use DGAs to randomly generate domain names that are registered on the go, making them harder to detect. It is therefore important to identify the domains generated by DGAs and block them before they can be used to establish communication between the bot and the C\&C center. There are several machine learning approaches proposed in the literature to address this issue including~\cite{antonakakis2012throw,schiavoni2014phoenix,schuppen2018fanci,Woodbridge2016, yu2017inline, rhodes2018, tran2018lstm, Saxe2017} and other work that we cite later in this paper. These well known state-of-the-art classifiers can be deployed in real-world DNS applications to detect DGA domain names and block them. While some work focuses on detecting DGAs from NXDomains~\cite{schuppen2018fanci}, our work aims to detect DGAs from traffic to domains that have already been resolved.

Commonly used approaches for DGA detection can be categorized according to how fast they are able to flag malicious activity in DNS traffic. As illustrated in \cref{TAB:RELATEDWORK}, some techniques work in a \textit{retrospective} manner, in which past DNS traffic, which is logged over a certain window, is analyzed in batches to detect anomalies. Other techniques work \textit{inline}, meaning that they can detect DGA domains as soon as they are queried. There are two ways in which inline DGA detection can happen:
\begin{itemize}[leftmargin=*]
    \item The domain first reaches the DGA classifier and if the classifier flags the domain as benign, then the query is passed to the DNS resolver to fetch the resolved IP address of the domain. However, if the classifier flags the domain as DGA, then the query will not be forwarded to the DNS resolver and it simply blocks the communication with that domain.
    \item The domain first queries through the DNS resolver; the DGA classifier uses the features learned from the DNS response to decide on whether the domain is DGA or not.
\end{itemize}
Our work fits into the second category of inline detection, where both the domain name and the side information features learned from the DNS query/response are used by the classifier for DGA detection. 
The side information features are carefully selected to allow \textit{inline DGA detection in the broader sense}. In the strict sense, inline DGA detection means that the information required to determine whether a domain name is DGA or not is available from the DNS query data alone. A DNS resolver can use the strictly inline DGA classifier's decision to determine if it is safe or not to resolve the query. A less conventional version of inline detection, which we refer to as ``inline DGA detection in the broader sense'', is one where data attributes from DNS responses are required (in addition to DNS queries). This means that the DNS resolver must resolve the query first, feed the information obtained to the DGA classifier, and then use the DGA classifier's decision to determine if it is safe to get the DNS response to the client or not. As we observe in our experimental results, taking information from DNS responses into account improves the ability of DGA classifiers to correctly detect DGA domains among resolvable traffic. We note that any dependence on data requiring queries to additional sources, such as the WHOIS database (as used for instance in \cite{curtin2019detecting, chin2018machine, li2019machine}), would disqualify the approach from inline detection, even in the broader sense.

Machine learning based approaches to detect DGA domain names in practice can also be categorized according to the information they leverage. One way is to train classifiers to detect DGA domain names using only the domain name string itself, see e.g.~\cite{schuppen2018fanci, Woodbridge2016, yu2017inline, tran2018lstm, choudhary2018algorithmically, sivaguru2018evaluation, Yu2018a}. The  alternative is to train the classifiers using context information such as the IP address of the domain, its geographic location, attributes from DNS response records etc.~in addition to the domain name \cite{schiavoni2014phoenix, curtin2019detecting, yadav2012detecting, bilge2014exposure, lison2017neural, li2019machine}. In our work, we combine both approaches. The advantage of the former approach is that it does not require gathering of additional information, which may be expensive to collect in real time, and that it allows the defenders to detect the DGA domain names and block them even before they can be resolved. The advantage of the latter approach is that side information is a lot harder for the attacker to manipulate than the domain name string itself, making machine learning models trained on side information potentially more robust against adversarial attacks. 



Adversarial machine learning is a research area focused on problems introduced by the use of machine learning techniques in adversarial environments in which an
intelligent adversary attempts to exploit the weaknesses in such techniques \cite{vorobeychik2018adversarial}. The \textit{adversarial attacks} of interest in this paper are \textit{evasion attacks} in which an adversary uses artificially crafted instances, called \textit{adversarial samples}, that are intentionally used to mislead a machine learning system and produce erroneous results. 
The goal of evasion attacks in the context of DGA detection is to generate domains that will be labeled as benign by the DGA classifier. The vulnerability of a classifier against evasion attacks is measured in terms of DGA detection rate, which is the proportion of the adversarial samples predicted as malicious by the classifier. Lower DGA detection rates indicate high vulnerability of the classifier to the attack. There exists several evasion attacks against DGA classifiers such as CharBot~\cite{peck2019charbot}, DeepDGA~\cite{anderson2016deepdga}, DeceptionDGA~\cite{spooren2019detection}, MaskDGA~\cite{sidi2019maskdga} and the DGAs (HMM \& PCFG-based) proposed by \cite{fu2017stealthy}. CharBot and MaskDGA are black-box targeted evasion attacks that do not require any knowledge about the DGA classifier and are intended to generate samples that can evade detection by \textit{any} classifier. On the other hand, DeceptionDGA is a white-box attack algorithm that uses the knowledge of features used by the DGA classifier to generate evading instances specific to a given classifier. Both types of attacks are found to be extremely powerful in generating domains that can evade detection by the DGA classifiers with high probability.


The main contributions of our work are:
\begin{itemize}[leftmargin=*]
    \item A comprehensive survey of lexical and side information features proposed in the literature on DGA detection.
    \item An experimental evaluation of the feasibility in collecting the features and their effectiveness when deployed for inline detection of DGAs in real streams of passive DNS traffic, which leads to a shortlist of features that are actually beneficial in practice.
    \item Experimental results that show how the side information features can make DGA classifiers more robust against adversarial attacks.
\end{itemize}
The paper is organized as follows. \Cref{SEC:RELATED} gives an overview of related work in the fields of adversarial machine learning and DGA detection. \Cref{SEC:FEATURES} provides a detailed overview of side information features that can be extracted from DNS traffic to aid in the detection of DGA domains. In \cref{sec:lexical}, we list the 26 human engineered lexical features that are extracted manually from the domain name string in order to train the RF classifier for DGA detection. \Cref{sec:classifiers} gives an overview of the different classifiers we will be studying and attempting to harden against adversarial attacks. \Cref{sec:experiments} describes the experimental setup and reports all of our empirical results. Finally, \cref{sec:conclusion} concludes the work.


%
%

\section{Related Work} \label{SEC:RELATED}
Given the importance of being able to detect and block DGA domain related traffic, it comes as no surprise that the problem of automatic DGA detection has received a considerable amount of attention over the last decade. There are various ways in which existing DGA detection approaches differ from each other. As illustrated in \cref{TAB:RELATEDWORK}, 
DGA detection can be categorized according to the kind of input that is required. Some techniques require just the \textit{domain name string}, while other techniques require \textit{side information}, or a combination of both. Both kinds of input have their own advantages and disadvantages. Methods that rely only on the domain name string are popular because side information is typically harder to obtain. On the other hand, features extracted from side information are harder to manipulate, making methods based on them more robust against adversarial attacks. All the approaches presented in our paper perform \textit{inline} DGA detection using domain name only, side information only and a combination of both domain name \& side information features.


Furthermore, the classifiers can be trained in two ways to detect if a given domain name is generated by a DGA or not. The first technique is the \textit{featureful approach}, where the classifier relies on human engineered features extracted from the domain names. The second technique is the \textit{featureless approach}, where the classifier learns the features automatically during the training process. Classifiers that are based on deep learning architectures like Long Short-Term Memory (LSTM) \cite{Woodbridge2016,tran2018lstm} and Convolutional Neural Network (CNN) models \cite{yu2017inline,Saxe2017} leverage the featureless approach, whereas models such as random forests (RFs) adopt the featureful approach. In our work, we will be using both featureful and featureless approaches to train random forest and deep learning classifiers for DGA detection.

\begin{table*}
\centering
\begin{tabular}{lll}
    \toprule
    Input & Retrospective & Inline \\
    \midrule
    \multirow{2}{*}{Domain name string}  & \multirow{2}{*}{\cite{antonakakis2012throw, pereira2018dictionary}} & \cite{tran2018lstm, Woodbridge2016, schuppen2018fanci, Yu2018a, joshi2019using}\\
    & & \cite{yu2019weakly, yu2017inline, sivaguru2018evaluation, choudhary2018algorithmically, wang2018detection} \\
    \midrule
    Side information features & \cite{kwon2016psybog, antonakakis2011detecting, yadav2012detecting, watkins2017using, khalil2016discovering} & our work\\
    \midrule
    Domain name string + & \cite{lison2017neural, ma2009beyond, schiavoni2014phoenix, singh2019detecting, abbink2017popularity} & \multirow{2}{*}{our work}\\
    side information features & \cite{bilge2014exposure, curtin2019detecting, chin2018machine, li2019machine} & \\
    \bottomrule
\end{tabular}
\caption{Overview of existing work on DGA detection \label{TAB:RELATEDWORK}
}
\end{table*}

\section{side information Features}\label{SEC:FEATURES}


\begin{figure}[ht]
    \centering
    \includegraphics[width=0.45\textwidth]{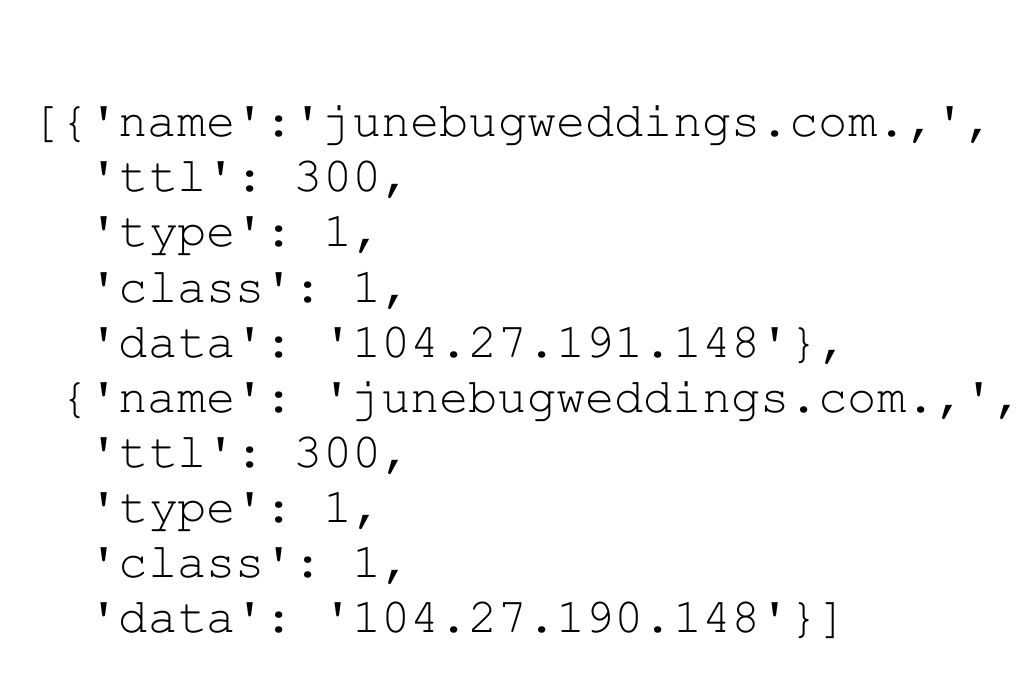}
    \caption{An example DNS resource record}
    \label{fig:resource-record}
\end{figure}

\begin{figure*}
    \centering
    \includegraphics[width=\textwidth]{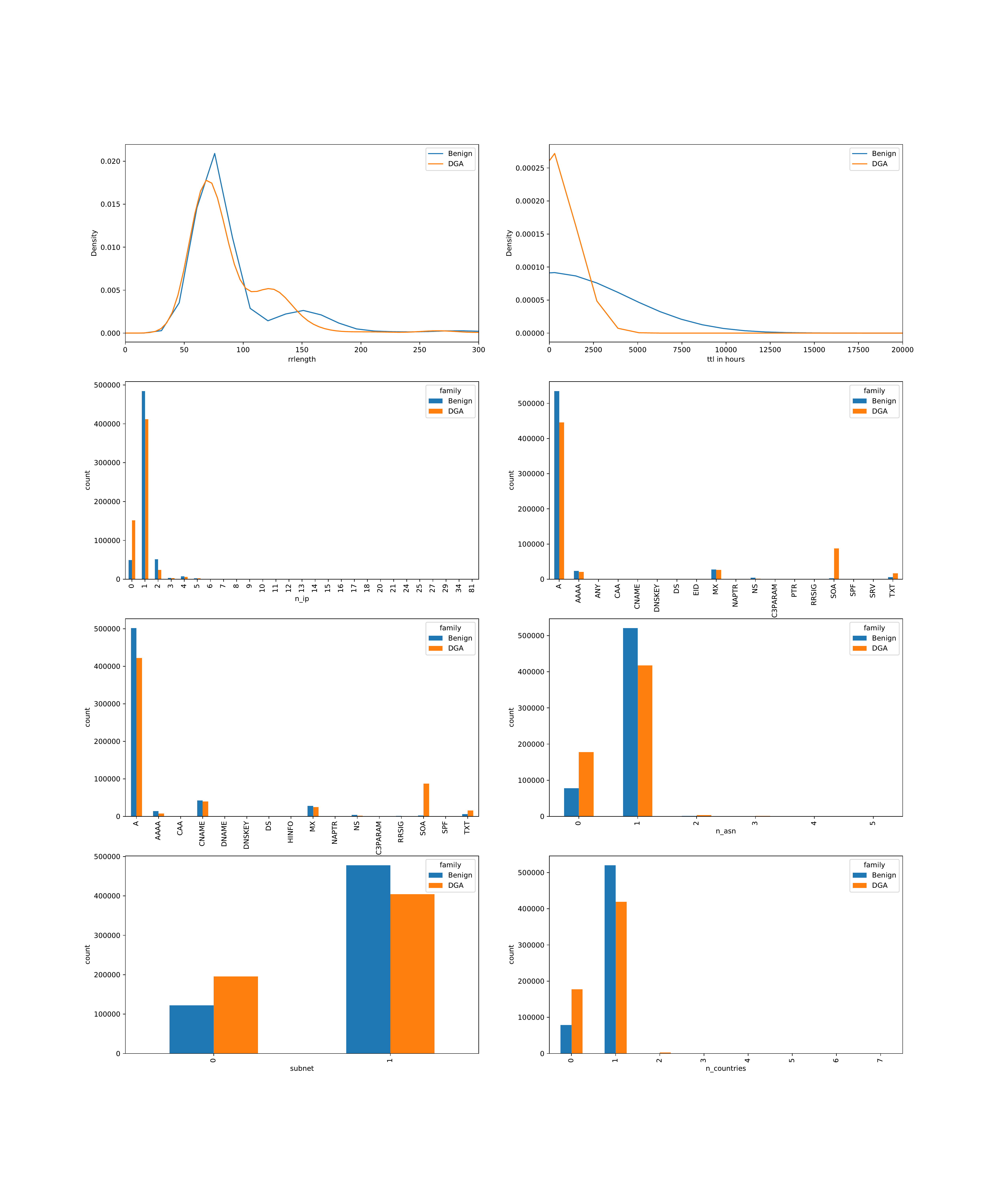}
    \caption{Comparison of values for side information features extracted from benign and DGA domains}
    \label{fig:dns-feature-analysis}
\end{figure*}

In this section we provide a detailed overview of side information features that can be extracted from DNS traffic to aid in the detection of DGA domains. An overview of all features is presented in \cref{tab:dns_features}, accompanied by a list of citations that illustrates the popularity of each kind of feature in the literature. The order of the side information features listed in \cref{tab:dns_features} indicates the importance of those features in DGA detection as ranked by the Random Forest model (see \cref{sec:classifiers}). Not all features are equally easy to obtain in practice, and their contribution to the predictive accuracy of DGA classifiers varies. The last column of \cref{tab:dns_features} indicates whether we retained the feature in our DGA-classifiers. \Cref{fig:resource-record} shows a sample resource record from which the side information features are extracted. In \cref{fig:resource-record}, the attribute ``name'' represents the fully qualified domain name (FQDN), ``ttl'' represents the time-to-live of the DNS query, ``type'' represents the resource record type, ``class'' represents the class of resource record and ``data'' represents the resolved IP address. Below we give a more in-depth description of each kind of feature, and its typical use in the the literature on DGA detection. \Cref{fig:dns-feature-analysis} shows a comparison of density plots for some of the side information features extracted from benign and DGA domain names, illustrating their predictive power. The different side information features are as follows:

\begin{sidewaystable*}
    \centering
    \begin{tabular}{lllc}
        \toprule
         Feature & Description & Reference & Retained \\
         \midrule
         rrlength & Resource record length & \cite{watkins2017using} & \checkmark\\
         country & Country name that the domain maps to & \cite{chin2018machine, li2019machine, lison2017neural, ma2009beyond} & \checkmark\\
         ttl & Time-to-live of the DNS query &  \cite{ma2009beyond, watkins2017using} & \checkmark\\
         n\_ip & Number of distinct IP addresses the domain maps to & \cite{bilge2014exposure, chin2018machine, li2019machine, lison2017neural} & \checkmark\\
         qtype & Type of DNS packet requested & \cite{watkins2017using} & \checkmark\\
         rtype & Record type of the DNS response & \cite{watkins2017using} & \checkmark\\
         n\_asn & Number of distinct ASNs the domain maps to & \cite{khalil2016discovering} & \checkmark\\
         subnet & Do all IPs belong to same subnet & \cite{li2019machine, chin2018machine} & \checkmark\\
         n\_countries & Number of distinct countries the domain maps to & \cite{bilge2014exposure, chin2018machine, li2019machine, lison2017neural, singh2019detecting} & \checkmark\\
         
         
         timestamp & Features derived from timestamp of the DNS query & \cite{kwon2016psybog, bilge2014exposure, lison2017neural} & \xmark\\
         opcode & Kind of DNS query & \cite{watkins2017using} & \xmark\\
         AA & Authoritative answer & \cite{watkins2017using} & \xmark\\
         QDCOUNT & Number of entries in question section & \cite{watkins2017using} & \xmark\\
         ANCOUNT & Number of resource records in answer section & \cite{watkins2017using} & \xmark\\
         NSCOUNT & Number of name servers in authoritative section & \cite{watkins2017using} & \xmark\\
         ARCOUNT & Number of resource records in additional record section & \cite{watkins2017using} & \xmark\\
         RCODE & Response code & \cite{watkins2017using} & \xmark\\
         rDNS & Reverse DNS query results & \cite{bilge2014exposure, chin2018machine, li2019machine} & \xmark\\
         TTL statistics & Mean, standard deviation etc. of time-to-live & \cite{bilge2014exposure, chin2018machine, li2019machine, lison2017neural} & \xmark\\
         n\_domains & Number of distinct domains associated with the IP & \cite{bilge2014exposure, chin2018machine, li2019machine, lison2017neural} & \xmark\\
         n\_queries & Number of queries for the domain and (domain, IP) pair & \cite{lison2017neural} & \xmark\\
         WHOIS features & Registrar, domain creation/expiration date etc. & \cite{curtin2019detecting, chin2018machine, li2019machine, ma2009beyond} & \xmark\\
         \bottomrule
    \end{tabular}
    \caption{side information features}
    \label{tab:dns_features}
\end{sidewaystable*}

\begin{itemize}[leftmargin=*]
    \item \textbf{rrlength:} This feature measures the length of the RData field, which is extracted directly from the DNS response resource record. The RData in a DNS response encompasses a list of resolved IP addresses, the time-to-live value of the query and the type of resource record.
    
    \item \textbf{country:} This feature refers to the geographic location that the resolved IP address maps to. If the DNS resource record contains multiple IP addresses, the country for each of the IP addresses is first identified. If all of the IP addresses belong to the same country, then this feature takes up that name. On the other hand, if any of the IP addresses map to a different location, then the value of this feature would be ``multi-valued''. Alternatively, if the location could not be identified, then this feature takes the value ``unknown''. This feature is then converted to categorical values that range between 0 and 185, which means that the domains in our data set map to 184 different countries plus the values ``multi-valued'' and ``unknown''.
    
    \item \textbf{ttl:} This feature represents the time-to-live value of the DNS query, which is the time interval that the resource record can be cached by the DNS resolver, and is directly obtained from the DNS response resource record.
    \cref{tab:ttl-values} compares the distribution of TTL values (in seconds), in terms of mean, standard deviation and median, for benign and DGA domains in our data set (see \cref{SUBSEC:DATA}). It can be seen that DGA domains are in general far more short-lived than benign domains. For better visibility in \cref{fig:dns-feature-analysis}, the density plot for TTL values are shown in hours instead of seconds.
    
    \begin{table}[ht]
        \centering
        \begin{tabular}{lrrr}
            \toprule
            Type & Mean TTL & SD TTL & Median TTL \\
            \midrule
            Benign & 109,447 & 1,421,829 & 3,600\\
            DGA & 29,255 & 4,701,205 & 900 \\
            \bottomrule
        \end{tabular}
        \caption{TTL distribution (in seconds) for benign and DGA domains}
        \label{tab:ttl-values}
    \end{table}
    
    \item \textbf{n\_ip:} This feature indicates the number of distinct IP addresses that are returned for the DNS domain lookup. It is manipulated directly by accessing the list of IPs contained in the RData field of the DNS response resource record.
    
    \item \textbf{qtype:} This feature represents the DNS query type that can be extracted from the question section of the DNS query. \Cref{fig:dns-feature-analysis} shows the different values for this features in our data set.
    
    \item \textbf{rtype:} This feature represents the resource record type that can be extracted directly from the RData field in the DNS response resource record. \Cref{fig:dns-feature-analysis} shows the different values for this features in our data set.
    
    
    \item \textbf{n\_asn:} This feature indicates the number of distinct autonomous system numbers that the IP addresses map to. The ASN for a given IP address is obtained by using Python Geolite2 Maxmind API.\footnote{\url{https://geoip2.readthedocs.io/en/latest/}}
    
    \item \textbf{subnet}: This feature is a boolean value that represents if all the IP addresses belong to the same subnet. A value of 0 indicates that one or more of the IP addresses, returned in the DNS response, belong to a different subnet and value of 1 indicates that all the IP addresses map to the same subnet.
    
    \item \textbf{n\_countries}: This feature represents the distinct number of countries that the resolved IP addresses map to. This feature has a very similar distribution when compared to the ``n\_asn'' feature, which can be observed in \cref{fig:dns-feature-analysis}.
    
    \item \textbf{timestamp:} The timestamp denotes the time at which the DNS query was issued by a host. This feature in itself may not be useful in detecting DGAs. Some of the past studies record all of the timestamps at which a particular domain name was queried and construct time-series data to analyze the periodicity at which the benign and DGA domains are queried \cite{bilge2014exposure, kwon2016psybog}, whereas \cite{lison2017neural} computes the lifespan of a domain by subtracting the first and last seen timestamps of the domain name. Such approaches require access to past DNS traffic and hence are regarded as ``retrospective''. Since we only focus on performing inline DGA detection in our work, we do not use the timestamp feature to perform DGA classification.
    
    \item \textbf{opcode:} This feature represents the kind of query such as standard query, inverse query, request for server status etc. In our data set, all the domains being queried belong to \textit{standard query} type and hence using this feature does not contribute in the prediction of DGA domain names.
    
    \item \textbf{aa:} This feature is a boolean flag which represents if the responding name server is an authority for the domain name being queried. The AA flag for all DNS responses in our data set has the same value ``True'' and hence we do not leverage the AA flag information while training our DGA classifiers.
    
    \item \textbf{qdcount, ancount, nscount, arcount:} At this time, our DNS traffic collector do not capture this information \& hence we do not use these features to train our model. However, it can be easily obtained from the DNS query and resource records.
    
    \item \textbf{rcode:} Since our data set comprises of resolved domain names only, the rcode remains ``0'' for all the samples and hence we discard this information.
    
    \item \textbf{TTL statistics:} This refers to a collection of features such as standard deviation, mean, minimum, maximum etc. of all time-to-live values extracted from the DNS response. While these features are relevant in a retrospective approach that investigates a domain based on \textit{all} DNS resource records related to it say during the past 24 hours, it is not meaningful for fast inline DGA detection. Indeed, since all of the TTL values in a single response record have constant values, it would not add value to include these statistics as features.
    
    \item \textbf{n\_domains:} This feature represents the number of distinct domain names that are mapped to a given IP address. In order to use this feature, one needs to maintain a bipartite graph that depicts the mapping for each (domain, IP) pair. Again, this method of performing graph inference is computationally intensive and does not contribute towards inline detection of DGA domains. Therefore we refrain from using this side information feature while training our DGA classifiers.
    
    \item \textbf{n\_queries:} Similar to ``timestamp'' and ``n\_domains'', this feature also requires storing and fetching of information from past DNS traffic and hence n\_queries cannot be used for inline detection of DGAs. 
    
    \item \textbf{WHOIS features:} Extracting WHOIS features such as registrar, domain creation/expiration date etc.~involves very expensive WHOIS queries. This affects the capability of the classifier to perform inline DGA detection on-the-go and hence we do not use any feature that require WHOIS queries.
    
\end{itemize}



%
%

\section{Lexical Features}\label{sec:lexical}

In this section we list the 26 human engineered lexical features that are extracted manually from the domain name string in order to train the RF classifier for DGA detection. \Cref{tab:features} shows a list of the lexical features used in B-RF and details on how the feature values are calculated are given below:


\begin{sidewaystable*}
    \centering
    \begin{tabular}{lllc}
        \toprule
        Feature & Description & Reference & Retained\\
        \midrule
        domain\_len & Domain name length & \cite{peck2019charbot, sivaguru2018evaluation, yu2017inline, choudhary2018algorithmically, schuppen2018fanci, joshi2019using, wang2018detection, chin2018machine, li2019machine} & \cmark\\
        sld\_len & Second level domain length & \cite{peck2019charbot, sivaguru2018evaluation, choudhary2018algorithmically} & \cmark\\
        tld\_len & Top level domain length & \cite{peck2019charbot, sivaguru2018evaluation, choudhary2018algorithmically} & \cmark\\
        uni\_domain & Domain Unique Characters length & \cite{peck2019charbot, sivaguru2018evaluation, choudhary2018algorithmically} & \cmark\\
        uni\_sld & SLD Unique Characters length & \cite{peck2019charbot, sivaguru2018evaluation, choudhary2018algorithmically} & \cmark\\
        uni\_tld & TLD Unique Characters length & \cite{peck2019charbot, sivaguru2018evaluation, choudhary2018algorithmically} & \cmark\\
        flag\_dga & Has malicious TLD & \cite{peck2019charbot, sivaguru2018evaluation, choudhary2018algorithmically, joshi2019using} & \cmark\\
        tld\_hash & TLD Hash & \cite{peck2019charbot, sivaguru2018evaluation, choudhary2018algorithmically, yu2017inline} & \cmark\\
        flag\_dig & Starts with Digit & \cite{peck2019charbot, sivaguru2018evaluation, choudhary2018algorithmically, yu2017inline} & \cmark\\
        sym & Symbol ratio & \cite{peck2019charbot, sivaguru2018evaluation, choudhary2018algorithmically, yu2017inline} & \cmark\\
        hex & Hex ratio & \cite{peck2019charbot, sivaguru2018evaluation, choudhary2018algorithmically, yu2017inline} & \cmark\\
        dig & Digit Ratio & \cite{peck2019charbot, sivaguru2018evaluation, choudhary2018algorithmically, schuppen2018fanci, wang2018detection, bilge2014exposure, chin2018machine, li2019machine} & \cmark\\
        vow & Vowel Ratio & \cite{peck2019charbot, sivaguru2018evaluation, choudhary2018algorithmically, yu2017inline, schuppen2018fanci, wang2018detection} & \cmark\\
        con & Consonant Ratio & \cite{peck2019charbot, sivaguru2018evaluation, choudhary2018algorithmically} & \cmark\\
        rep\_char\_ratio & Ratio of Repeated Characters & \cite{peck2019charbot, sivaguru2018evaluation, schuppen2018fanci} & \cmark\\
        cons\_con\_ratio & Ratio of Consecutive Consonants & \cite{peck2019charbot, sivaguru2018evaluation, schuppen2018fanci, wang2018detection} & \cmark\\
        cons\_dig\_ratio & Ratio of Consecutive Digits & \cite{peck2019charbot, sivaguru2018evaluation, schuppen2018fanci} & \cmark\\
        tokens\_sld & Number of tokens in SLD & \cite{peck2019charbot, sivaguru2018evaluation, choudhary2018algorithmically, joshi2019using} & \cmark\\
        digits\_sld & Number of digits in SLD & \cite{peck2019charbot, sivaguru2018evaluation, choudhary2018algorithmically, joshi2019using} & \cmark\\
        ent & Entropy of characters in SLD & \cite{peck2019charbot, sivaguru2018evaluation, choudhary2018algorithmically, yu2017inline, schuppen2018fanci, wang2018detection} & \cmark\\
        gni & Gini Index of characters in SLD & \cite{peck2019charbot, sivaguru2018evaluation, choudhary2018algorithmically, yu2017inline} & \cmark\\
        cer & Classification error of characters in SLD & \cite{peck2019charbot, sivaguru2018evaluation, choudhary2018algorithmically, yu2017inline} & \cmark \\
        2gram\_med & 2-Gram Median of characters in SLD & \cite{peck2019charbot, sivaguru2018evaluation, choudhary2018algorithmically, yu2017inline} & \cmark \\
        3gram\_med & 3-Gram Median of characters in SLD & \cite{peck2019charbot, sivaguru2018evaluation, choudhary2018algorithmically, yu2017inline} & \cmark \\
        2gram\_cmed & 2-Gram Circle Median of characters in SLD & \cite{peck2019charbot, sivaguru2018evaluation, choudhary2018algorithmically} & \cmark\\
        3gram\_cmed & 3-Gram Circle Median of characters in SLD & \cite{peck2019charbot, sivaguru2018evaluation, choudhary2018algorithmically} & \cmark\\
        \bottomrule
    \end{tabular}
    \caption{Lexical features used by B-RF}
    \label{tab:features}
\end{sidewaystable*}

\begin{itemize}[leftmargin=*]
    \item \textbf{domain\_len:} This feature represents the length of the domain name, which is the number of characters in the SLD.TLD pair. For example, we refer ``google.com'' as the domain name, where ``google'' indicates the SLD (second level domain) and ``com'' indicates the TLD (top level domain). The value of the feature domain\_len for the domain name ``google.com'' is 10.
    
    \item \textbf{sld\_len:} This feature represents the number of characters in the second level domain.
    
    \item \textbf{tld\_len:} This feature represents the number of characters in the top level domain.
    
    \item \textbf{uni\_domain:} This feature represents the number of unique characters in the domain name, after removing special characters such as `.' \& `-' from the domain name.
    
    \item \textbf{uni\_sld:} This feature represents the number of unique characters in the second level domain, after removing special characters such as `.' \& `-' from the SLD.
    
    \item \textbf{uni\_tld:} This feature represents the number of unique characters in the top level domain, after removing special characters such as `.' \& `-' from the TLD.
    
    \item \textbf{flag\_dga:} This feature represents a boolean value (0 or 1) that indicates if the domain name contains any of the following TLDs, which are known to be frequently associated with malicious activities\footnote{\url{https://www.spamhaus.org/statistics/tlds/}}: ``study'', ``party'', ``click'', ``top'', ``gdn'', ``gq'', ``asia'', ``cricket'', ``biz'', ``cf''.
    
    \item \textbf{tld\_hash:} This feature represents the hash value of top level domain.
    
    \item \textbf{flag\_dig:} This feature represents a boolean value that indicates if the domain name starts with a digit/number (0-9).
    
    \item \textbf{sym:} This feature represents the ratio of number of special characters in the SLD to the total number of characters in SLD (sld\_len).
    
    \item \textbf{hex:} This feature represents the ratio of number of hexadecimal characters (0-9 \& a-f) in the SLD to the total number of characters in the SLD.
    
    \item \textbf{dig:} This feature represents the ratio of number of digits (0-9) in the SLD to the total number of characters in the SLD.
    
    \item \textbf{vow:} This feature represents the ratio of number of vowels (`a', `e', `i', `o', `u') in the SLD to the total number of characters in the SLD.
    
    \item \textbf{con:} This feature represents the ratio of number of consonants in the SLD to the total number of characters in the SLD.
    
    \item \textbf{rep\_char\_ratio:} This feature represents the ratio of number of characters that occurs more than once in the SLD to the total number of unique characters in the SLD.
    
    \item \textbf{cons\_con\_ratio:} This feature represents the ratio of consecutive consonants (such as ``ct'', ``fk'', ``ns'' etc.) to the length of the domain (domain\_len).
    
    \item \textbf{cons\_dig\_ratio:} This feature represents the ratio of consecutive digits (such as ``92'', ``24'', ``75'' etc.) to the length of the domain (domain\_len).
    
    \item \textbf{tokens\_sld:} This feature represents the number of tokens in the SLD, where a token indicates sequence of characters separated by `-'.
    
    \item \textbf{digits\_sld:} This feature represents the total number of digits in the SLD.
    
    \item \textbf{ent:} This feature represents the normalized entropy value of the characters in SLD and is calculated using the formula:
    $$ \mbox{ent} = \frac{\sum_{i=1}^{n} p_i*\log_2(p_i)}{\log_2(\mbox{sld\_len})} $$
    where $n$ represents the number of unique characters in the SLD and $p_{i}$ represents the proportion between the frequency of the unique character $c_{i}$ in the SLD to the total number of  unique characters in the SLD.
    
    \item \textbf{gni:} This feature represents the Gini value of the characters in SLD and is calculated using the formula:
    $$ \mbox{gni} = 1 - \sum_{i=1}^{n} p_i^2 $$
    where $n$ represents the number of unique characters in the SLD and $p_{i}$ represents the proportion between the frequency of the unique character $c_{i}$ in the SLD to the total number of  unique characters in the SLD.
    
    \item \textbf{cer:} This feature represents the classification of error of characters in SLD, which is computed using the formula:
    $$ \mbox{cer} = 1 - \max_{i = 1, \dots, n}~p_i $$
    where $p_{i}$ represents the proportion between the frequency of the unique character $c_{i}$ in the SLD to the total number of  unique characters in the SLD.
    
    \item \textbf{2gram\_med:} This feature represents the median of 2-gram frequencies in SLD.
    
    \item \textbf{3gram\_med:} This feature represents the median of 3-gram frequencies in SLD.
    
    \item \textbf{2gram\_cmed:} In order to compute this feature, the SLD of the domain is concatenated again with the SLD. (i.e) For example, if ``google'' is the SLD, a string such as ``googlegoogle'' is formed. The 2gram\_med is then calculated on this newly formed string ``googlegoogle'' to obtain the value of this feature.
    
     \item \textbf{3gram\_cmed:} In order to compute this feature, the SLD of the domain is concatenated again with the SLD. (i.e) For example, if ``yahoo'' is the SLD, a string such as ``yahooyahoo'' is formed. The 3gram\_med is then calculated on this newly formed string ``yahooyahoo'' to obtain the value of this feature.
    
\end{itemize}

%
%

\section{DGA Classifiers}\label{sec:classifiers}

We consider three different DGA classifiers in this work, which we detail below. We chose one model representative of the featureful approach (B-RF), one deep learning model which represents the featureless approach (LSTM.MI) and finally a hybrid model which combines both approaches (LSTM.MI+B-RF).

\subsection{B-RF}
B-RF is a DGA classifier based on random forests. It consists of 100 trees and each tree is trained using a subset of the feature space to avoid overfitting. Entropy is used as the criterion to decide the split attribute while growing the trees in the random forest. There are 3 variants of B-RF classifier, each trained either on lexical features (as the RF classifier in \cite{sivaguru2018evaluation}) or DNS features, or a combination of both lexical and DNS features. The performance of these variants of the B-RF classifier is listed in the first 3 rows of \cref{tab:results}.

\subsection{LSTM.MI}
Woodbridge et al.~\cite{Woodbridge2016} were the first to propose deep learning for DGA domain name detection. Their DGA classifier is a neural network consisting of an embedding layer, an LSTM layer, and a single node output layer with sigmoid activation. In this paper, we use the LSTM.MI model that was proposed recently by Tran et al.~\cite{tran2018lstm}. Its architecture is very similar to that of Woodbridge et al.~\cite{Woodbridge2016}; the main distinction is that the LSTM.MI model is trained with a cost-sensitive learning algorithm that takes class imbalances into account. This allows the LSTM.MI approach to achieve slightly better results than the original LSTM approach (see \cite{tran2018lstm,sivaguru2018evaluation}). 
The 4th row in \cref{tab:results} shows the performance of the LSTM.MI classifier. It operates directly on the domain name string, instead of on lexical features extracted from it. Characters in the domain name are converted to lower case and are encoded with categorical values, ranging from 1 to 38, to represent `.', `-', digits from 0 to 9 \& characters from `a' to `z'. All the domains in our data are fixed to a length of 77 characters, which is the length of the longest domain name in our data set. Domains that are shorter than 77 characters are padded with zeroes in the left.




\subsection{LSTM.MI+B-RF}

\begin{figure*}
    \centering
    \includegraphics[scale=0.7]{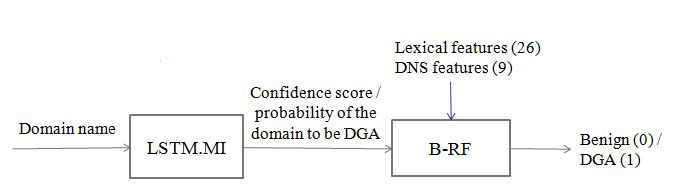}
    \caption{DGA detection using LSTM.MI + RF model}
    \label{fig:lstm-rf}
\end{figure*}

The hybrid LSTM.MI+B-RF classifier
combines both LSTM.MI and B-RF architectures by training a B-RF classifier with features listed in \cref{tab:dns_features,tab:features}, in addition to the confidence score obtained from the LSTM.MI model for that domain name. The confidence score ranges between 0 and 1, signifying the probability of the domain being a DGA as predicted by the LSTM.MI classifier. The above workflow of DGA detection using LSTM.MI+B-RF setup is depicted in \cref{fig:lstm-rf}. The last two rows in \cref{tab:results} represent the performance of this DGA classifier.

%
%

\section{Experimental results}\label{sec:experiments}

\subsection{Dataset}\label{SUBSEC:DATA}

In the first experiment, we train and evaluate the DGA classifiers from \cref{sec:classifiers} on a dataset with 600,000 DGAs (positive) and 600,000 benign (negative) samples.
\Cref{tab:examples} shows some examples of DGA \& benign domains. The training data points originate from a real-time stream of passive DNS data, consisting of roughly 10-12 billion DNS queries per day collected from subscribers including ISPs (Internet Service Providers), schools, and businesses. From this traffic, the positive samples are collected by retaining resolved domain names that are listed in DGArchive\footnote{\url{https://dgarchive.caad.fkie.fraunhofer.de/}}, a blacklist containing known DGA domains \cite{plohmann2016comprehensive}. Dictionary DGAs, which are human-readable DGA domains belonging to malware families such as suppobox, gozi, matsnu and nymaim2 are discarded from the training set. This is because these DGAs look more like benign domains and confuse the DGA classifiers~\cite{pereira2018dictionary}. Since this work is primarily aimed at measuring the impact of adversarial instances such as CharBot, we exclude samples from Dictionary DGAs.
The benign samples are collected based on a predefined set of heuristics as listed below: 

\begin{itemize}[leftmargin=*]
    \item Domain name should have valid DNS characters only (digits, letters, dot and hyphen)
    \item Domain has to be resolved at least once for every day between June 01, 2019 and July 31, 2019.
    \item Domain name should have a valid public suffix
    \item Characters in the domain name are not all digits (after removing `.' and `-')
    \item Domain should have at most four labels (Labels are sequence of characters separated by a dot)
    \item Length of the domain name is at most 255 characters
    \item Longest label is between 7 and 64 characters
    \item Longest label is more than twice the length of the TLD
    \item Longest label is more than 70\% of the combined length of all labels
    \item Excludes IDN (International Distribution Network) domains (such as domains starting with \texttt{xn-{}-})
    \item Domain must not exist in DGArchive
\end{itemize}

Both the DGA and benign domains in the data set are collected from real-time passive DNS traffic that was observed in February 2019. The domains in the data set are then preprocessed by following the two steps mentioned below:

\begin{itemize}[leftmargin=*]
    \item Retain only the SLD \& TLD of the domain name and discard any 3LD (third level domain) or any other label if present. For example, for the domain name ``www.google.com'', the 3LD which is ``www'' is removed and the SLD.TLD which ``google.com'' is retained.
    \item All the alphabetical characters in the domain name are converted to its corresponding lower case characters.
\end{itemize}

\begin{table*}
    \centering
    \begin{tabular}{ll}
         \toprule
         Benign domains (labeled 0) & DGA domains (labeled 1) \\
         \midrule
         7ft4.com & vocom.eu \\
         sgtobel.ch & leadhelp.net \\
         intimvoronezh.net & 1b6a95e6b5d4.com \\
         essc-tabriz.com & korpncyeajsgeatkopoqs.info \\
         konsaltbezopasnost.ru & kndydusmrlrofrcmfuayfmswrkytl.biz \\
         \bottomrule
    \end{tabular}
    \caption{Some examples fo benign vs DGA domain names}
    \label{tab:examples}
\end{table*}

\subsection{Performance evaluation of DGA classifiers}
The true positive rate (TPR) and false positive rate (FPR) for the DGA classifiers are calculated as follows:
\begin{align*}
    \mbox{TPR} &= \frac{\mbox{TP}}{\mbox{TP}+\mbox{FN}}, &
    \mbox{FPR} &= \frac{\mbox{FP}}{\mbox{FP}+\mbox{TN}}
\end{align*}
where TP, TN, FP \& FN represent the number of true positives, true negatives, false positives and false negatives respectively. The predictive performance of the classifiers is evaluated using 5-fold cross-validation for metrics such as TPR and Area Under the Receiver Operating Characteristic (ROC) Curve (AUC) as tabulated in \cref{tab:results}. In cybersecurity applications, it is important to achieve high TPR for a very low FPR. This is because it is undesirable to block a large number of benign domains in real-world traffic as this interferes with users' legitimate business. Hence all the reported metrics are thresholded at a very low FPR of 0.1\%. We also obtain the ROC curve by plotting the TPR against the FPR of the classifiers and the AUC is subsequently obtained by taking the integral of the ROC curve. The AUC is a measure of how well the trained classifier can distinguish between the classes. Specifically, it can be interpreted as the probability that the classifier will output a higher score for a randomly chosen DGA domain than it would for a randomly chosen benign domain. An ideal classifier has an AUC score of 1, indicating it will always rank DGA domains higher than benign domains. This makes it possible to use the classifier to perfectly separate the classes via an appropriate threshold on its output scores. A classifier that just randomly guesses the outcome achieves an AUC of 0.5 and a classifier with AUC 0 has basically \textit{inverted} all predictions, i.e. samples labeled as 0 are predicted as 1 by the classifier and vice versa. In addition to the AUC score, the AUC at a fixed FPR of 0.1\% is also reported. This thresholded AUC represents the integral of the ROC curve for a FPR of 0 to 0.001.

\begin{table*}[ht]
    \centering
    \begin{tabular}{llcc}
        \toprule
        \multirow{3}{*}{Model} & \multirow{3}{*}{Features} & \multicolumn{2}{c}{Performance metrics} \\
        \cmidrule{3-4}
        & & AUC@ & TPR@ \\
        & & 0.1\%FPR & 0.1\%FPR \\
        \midrule
        \multirow{3}{*}{B-RF} & DNS & 53.23\% & 16.21\% \\
        & Lexical & 89.78\% & 97.44\% \\
        & DNS + Lexical & \textbf{98.19\%} & \textbf{99.42\%} \\
        \midrule
        LSTM.MI & Domain name string & 94.47\% & 98.80\% \\
        \midrule
        \multirow{2}{*}{LSTM.MI + B-RF} & Domain name string + DNS & 96.51\% & 99.89\% \\
        & Domain name string + DNS + Lexical & \textbf{99.17\%} & \textbf{99.91\%} \\
        \bottomrule
    \end{tabular}
    \caption{Performance evaluation of DGA classifiers using 5-fold cross-validation}
    \label{tab:results}
\end{table*}

There are several interesting observations to be made based on \cref{tab:results}. First, looking at the AUC@1\%FPR column, one can see that the predictive performance for inline DGA detection based on DNS features alone does not perform well: the B-RF/DNS based model achieves an AUC@1\%FPR of only 53.23\%. Second, when it comes to DGA detection based on the domain string alone, the deep learning approach (LSTM.MI) clearly outperforms the random forest approach (B-RF/Lexical) at 94.47\% vs.~89.78\%. This is fully in line with previous findings \cite{tran2018lstm,yu2019weakly}. Third, the most interesting and novel result from \cref{tab:results}
is that the DGA classifiers, when trained with \textit{both} lexical and side information features, have the best overall performance in terms of AUC score and TPR, namely 99.17\% for the architecture from \cref{fig:lstm-rf}.


\subsection{Real Traffic Analysis}
Next, we apply the best performing classifiers in \cref{tab:results} on one day of real traffic DNS traffic to evaluate their predictive performance in real-time. We collected a set of resolved domains that were observed on March 26, 2019 to perform this analysis. As part of pre-processing, the fully qualified domain names are validated against the heuristics mentioned in \cref{SUBSEC:DATA}, in order to maintain consistency with the training data set. The domains that satisfy the heuristics are then retained in this experiment after discarding the third level domain (3LD/subdomain) from the domain name, if present. This resulted in a set consisting of 66,440,681 domains (contains duplicate domains with SLD.TLD pairs), out of which 1,159,662 domains were found in DGArchive and 14,653,217 domains were found in Alexa. There is also an overlap of 1,124,467 domains between the Alexa whitelist and DGArchive blacklist.

\Cref{tab:real-traffic-analysis} shows a comparison of the number of domains that were flagged as DGA by the LSTM.MI, B-RF and LSTM.MI+B-RF classifier. The B-RF model (in \cref{tab:real-traffic-analysis}) has the highest true positive rate among the 3 models being compared. Out of the 1.87M domains flagged as DGA by the classifier, approximately 61\% were found in DGArchive. Although the LSTM.MI classifier catches the highest number of DGAs in real-traffic, the true positive rate is 34\% which is 27\% lower than the B-RF classifier. However, as seen in the last row of \cref{tab:real-traffic-analysis}, the B-RF also has the highest number of false positives. This could be due to the fact that there is a large number of overlapping domains between Alexa and DGArchive as mentioned earlier in this section. A good workaround to reduce the number of false positives during the deployment is to check the flagged domains against Alexa before making the final decision.

\begin{table*}[ht]
    \centering
    \begin{tabular}{lccc}
         \toprule
         Model & LSTM.MI & B-RF & LSTM.MI+B-RF\\
         \midrule
         \multirow{2}{*}{Features} & \multirow{2}{*}{Domain name} & \multirow{2}{*}{DNS + Lexical} & Domain name + \\
         & & & DNS + Lexical\\
         \midrule
         Out of the $\sim$66M domains in real-traffic, & \multirow{3}{*}{3,400,017} & \multirow{3}{*}{1,877,784} & \multirow{3}{*}{2,170,056} \\
         number of domains flagged as DGA & & & \\
         by the classifier & & & \\
         \midrule
         Out of the domains flagged as DGA & \multirow{3}{*}{1,151,750} & \multirow{3}{*}{1,149,689} & \multirow{3}{*}{1,150,116}\\
         by the classifier, & & & \\
         number of domains found in DGArchive & & & \\
         \midrule
         Out of the domains flagged as DGA & \multirow{3}{*}{1,626,232} & \multirow{3}{*}{1,717,638} & \multirow{3}{*}{1,420,319}\\
         by the classifier, & & & \\
         number of domains found in Alexa & & & \\
         \bottomrule
    \end{tabular}
    \caption{Real traffic analysis of DGA classifiers on 66,440,662 ($\sim$66M) domains}
    \label{tab:real-traffic-analysis}
\end{table*}


\subsection{Defense against Adversarial ML}
The use of side information is important in the context of adversarial ML because  side information is a lot harder to manipulate than the domain name string itself \cite{curtin2019detecting}. 
In order to test this, we generated 1,000 DGA domains with CharBot~\cite{peck2019charbot}, a simple DGA algorithm that was written specifically to evade existing DGA classifiers. Since, to the best of our knowledge, CharBot has not been deployed yet in the wild, we cannot collect side information for CharBot domains from real traffic. 
Instead, we pair up the CharBot domains with the DNS features obtained from 1,000 randomly sampled DGA domains in real traffic. To avoid any bias in the selection of DNS features for CharBot domains, we perform the random sampling for 5 trials and create 5 sets of CharBot DNS features. The lexical features extracted for CharBot are appended with the DNS features, which can then be exposed to DGA classifiers for detection of malicious domains. The idea here is to test if the DGA classifiers trained on side information features are successful in detecting CharBot domains. 

\Cref{tab:CharBot-detection} shows the CharBot detection rate, which is the average proportion of CharBot domains that were flagged as DGA by the classifiers over the 5 randomized trials. Higher values of CharBot detection rate indicates that the classifier is more robust against new DGAs or adversarial attacks. As expected, the B-RF model trained on both lexical and side information features detects 20\% of CharBot domains as DGA/malicious, which is 12\% more than the LSTM.MI model. This clearly indicates that the use of side information features to train the DGA classifier makes it more robust against adversarial samples like CharBot domains, when compared to classifiers that rely only on the domain name for DGA detection.


\begin{table*}
    \centering
    \begin{tabular}{llr}
         \toprule
         Classifier & Features & DGA (CharBot) detection rate \\
         \midrule
         \multirow{3}{*}{B-RF} & DNS & \SI{1.70}{\percent} \textpm\ \SI{0.24}{\percent}\\
         & Lexical & \SI{3.80}{\percent} \textpm\ \SI{0.0}{\percent}\\
         & Lexical + DNS & \SI{20.06}{\percent} \textpm\ \SI{0.56}{\percent}\\
         \midrule
         LSTM.MI & Domain name string & \SI{8.00}{\percent} \textpm\ \SI{0.0}{\percent}\\
         \midrule
         \multirow{2}{*}{LSTM.MI+B-RF} & Domain name string + DNS & \SI{14.98}{\percent} \textpm\ \SI{0.63}{\percent}\\
         & Domain name string + Lexical + DNS & \SI{15.76}{\percent} \textpm\ \SI{0.53}{\percent} \\
         \bottomrule
    \end{tabular}
    \caption{Detection rate of CharBot domains as DGA}
    \label{tab:CharBot-detection}
\end{table*}

\section{Conclusion}\label{sec:conclusion}
In this paper, we proposed and evaluated state-of-the-art classifiers for inline DGA detection using side information features that are easily obtained from DNS query and response. Results from \cref{tab:results,tab:CharBot-detection} show that using side information in addition to the domain name to train classifiers not only improves the predictive performance, but also makes it more robust against adversaries like CharBot, when compared to the classifiers that use just the domain name to detect DGAs. 
Additionally, the side information features in our approach are carefully chosen to perform lightweight \textit{inline} detection of DGA domains, and do not rely on external sources such as WHOIS for feature extraction.\\

\noindent
\textbf{Acknowledgement.} We gratefully acknowledge the support of NVIDIA Corporation with the donation of the Titan Xp GPU used for this research. Jonathan Peck is sponsored by a fellowship of the Research Foundation Flanders (FWO).

\balance
\bibliographystyle{IEEEtran}
\bibliography{main}

\begin{thebibliography}{10}
\providecommand{\url}[1]{#1}
\csname url@samestyle\endcsname
\providecommand{\newblock}{\relax}
\providecommand{\bibinfo}[2]{#2}
\providecommand{\BIBentrySTDinterwordspacing}{\spaceskip=0pt\relax}
\providecommand{\BIBentryALTinterwordstretchfactor}{4}
\providecommand{\BIBentryALTinterwordspacing}{\spaceskip=\fontdimen2\font plus
\BIBentryALTinterwordstretchfactor\fontdimen3\font minus
  \fontdimen4\font\relax}
\providecommand{\BIBforeignlanguage}[2]{{%
\expandafter\ifx\csname l@#1\endcsname\relax
\typeout{** WARNING: IEEEtran.bst: No hyphenation pattern has been}%
\typeout{** loaded for the language `#1'. Using the pattern for}%
\typeout{** the default language instead.}%
\else
\language=\csname l@#1\endcsname
\fi
#2}}
\providecommand{\BIBdecl}{\relax}
\BIBdecl

\bibitem{plohmann2016comprehensive}
D.~Plohmann, K.~Yakdan, M.~Klatt, J.~Bader, and E.~Gerhards-Padilla, ``{A
  Comprehensive Measurement Study of Domain Generating Malware},'' in
  \emph{25th {USENIX} Security Symposium}, 2016, pp. 263--278.

\bibitem{antonakakis2012throw}
M.~Antonakakis, R.~Perdisci, Y.~Nadji, N.~Vasiloglou, S.~Abu-Nimeh, W.~Lee, and
  D.~Dagon, ``{From Throw-Away Traffic to Bots: Detecting the Rise of DGA-Based
  Malware.}'' in \emph{{USENIX} Security Symposium}, vol.~12, 2012, pp.
  491--506.

\bibitem{schiavoni2014phoenix}
S.~Schiavoni, F.~Maggi, L.~Cavallaro, and S.~Zanero, ``{Phoenix: DGA-based
  Botnet Tracking and Intelligence},'' in \emph{International Conference on
  Detection of Intrusions and Malware, and Vulnerability Assessment}.\hskip 1em
  plus 0.5em minus 0.4em\relax Springer, 2014, pp. 192--211.

\bibitem{schuppen2018fanci}
S.~Sch{\"u}ppen, D.~Teubert, P.~Herrmann, and U.~Meyer, ``{FANCI: Feature-based
  Automated NXDomain Classification and Intelligence},'' in \emph{27th {USENIX}
  Security Symposium}, 2018, pp. 1165--1181.

\bibitem{Woodbridge2016}
J.~Woodbridge, H.~S. Anderson, A.~Ahuja, and D.~Grant, ``{Predicting Domain
  Generation Algorithms with Long Short-Term Memory Networks},'' \emph{preprint
  arXiv:1611.00791}, 2016.

\bibitem{yu2017inline}
B.~Yu, D.~L. Gray, J.~Pan, M.~De~Cock, and A.~C. Nascimento, ``{Inline DGA
  Detection with Deep Networks},'' in \emph{2017 IEEE International Conference
  on Data Mining Workshops (ICDMW)}, 2017, pp. 683--692.

\bibitem{rhodes2018}
J.~Koh and B.~Rhodes, ``{Inline Detection of Domain Generation Algorithms with
  Context-Sensitive Word Embeddings},'' in \emph{Proceedings of 2018 IEEE
  International Conference on Big Data}, 2018, pp. 2965--2970.

\bibitem{tran2018lstm}
D.~Tran, H.~Mac, V.~Tong, H.~A. Tran, and L.~G. Nguyen, ``{A LSTM based
  framework for handling multiclass imbalance in DGA botnet detection},''
  \emph{Neurocomputing}, vol. 275, pp. 2401--2413, 2018.

\bibitem{Saxe2017}
J.~Saxe and K.~Berlin, ``{eXpose: A Character-Level Convolutional Neural
  Network with Embeddings For Detecting Malicious URLs, File Paths and Registry
  Keys},'' \emph{preprint arXiv:1702.08568}, 2017.

\bibitem{curtin2019detecting}
R.~R. Curtin, A.~B. Gardner, S.~Grzonkowski, A.~Kleymenov, and A.~Mosquera,
  ``{Detecting DGA Domains with Recurrent Neural Networks and Side
  Information},'' in \emph{Proceedings of the 14th International Conference on
  Availability, Reliability and Security}.\hskip 1em plus 0.5em minus
  0.4em\relax ACM, 2019.

\bibitem{chin2018machine}
T.~Chin, K.~Xiong, C.~Hu, and Y.~Li, ``{A Machine Learning Framework for
  Studying Domain Generation Algorithm DGA-based Malware},'' in
  \emph{International Conference on Security and Privacy in Communication
  Systems}.\hskip 1em plus 0.5em minus 0.4em\relax Springer, 2018, pp.
  433--448.

\bibitem{li2019machine}
Y.~Li, K.~Xiong, T.~Chin, and C.~Hu, ``{A Machine Learning Framework for Domain
  Generation Algorithm DGA-Based Malware Detection},'' \emph{IEEE Access},
  vol.~7, pp. 32\,765--32\,782, 2019.

\bibitem{choudhary2018algorithmically}
C.~Choudhary, R.~Sivaguru, M.~Pereira, B.~Yu, A.~C. Nascimento, and M.~De~Cock,
  ``{Algorithmically Generated Domain Detection and Malware Family
  classification},'' in \emph{International Symposium on Security in Computing
  and Communication}.\hskip 1em plus 0.5em minus 0.4em\relax Springer, 2018,
  pp. 640--655.

\bibitem{sivaguru2018evaluation}
R.~Sivaguru, C.~Choudhary, B.~Yu, V.~Tymchenko, A.~Nascimento, and M.~De~Cock,
  ``{An Evaluation of DGA Classifiers},'' in \emph{2018 IEEE International
  Conference on Big Data}, 2018, pp. 5058--5067.

\bibitem{Yu2018a}
B.~Yu, J.~Pan, J.~Hu, A.~Nascimento, and M.~{De Cock}, ``{Character Level Based
  Detection of DGA Domain Names},'' in \emph{Proc. WCCI}, 2018, pp. 4168--4175.

\bibitem{yadav2012detecting}
S.~Yadav, A.~K.~K. Reddy, A.~N. Reddy, and S.~Ranjan, ``{Detecting
  Algorithmically Generated Domain-Flux Attacks with DNS Traffic Analysis},''
  \emph{IEEE/ACM Transactions on Networking}, vol.~20, no.~5, pp. 1663--1677,
  2012.

\bibitem{bilge2014exposure}
L.~Bilge, S.~Sen, D.~Balzarotti, E.~Kirda, and C.~Kruegel, ``{Exposure: A
  Passive DNS Analysis Service to Detect and Report Malicious Domains},''
  \emph{ACM Transactions on Information and System Security (TISSEC)}, vol.~16,
  no.~4, 2014.

\bibitem{lison2017neural}
P.~Lison and V.~Mavroeidis, ``{Neural Reputation Models learned from Passive
  DNS Data},'' in \emph{2017 IEEE International Conference on Big Data}, 2017,
  pp. 3662--3671.

\bibitem{vorobeychik2018adversarial}
Y.~Vorobeychik and M.~Kantarcioglu, ``{Adversarial Machine Learning},''
  \emph{Synthesis Lectures on Artificial Intelligence and Machine Learning},
  vol.~12, no.~3, pp. 1--169, 2018.

\bibitem{peck2019charbot}
J.~Peck, C.~Nie, R.~Sivaguru, C.~Grumer, F.~Olumofin, B.~Yu, A.~Nascimento, and
  M.~De~Cock, ``{CharBot: A Simple and Effective Method for Evading DGA
  Classifiers},'' \emph{IEEE Access}, vol.~7, pp. 91\,759--91\,771, 2019.

\bibitem{anderson2016deepdga}
H.~S. Anderson, J.~Woodbridge, and B.~Filar, ``{DeepDGA: Adversarially-Tuned
  Domain Generation and Detection},'' in \emph{Proceedings of the 2016 ACM
  Workshop on Artificial Intelligence and Security}, 2016, pp. 13--21.

\bibitem{spooren2019detection}
J.~Spooren, D.~Preuveneers, L.~Desmet, P.~Janssen, and W.~Joosen, ``{Detection
  of Algorithmically Generated Domain Names used by Botnets: A Dual Arms
  Race.}'' in \emph{Proceedings of the 34th ACM/SIGAPP Symposium On Applied
  Computing}.\hskip 1em plus 0.5em minus 0.4em\relax Association for Computing
  Machinery, 2019, pp. 1902--1910.

\bibitem{sidi2019maskdga}
L.~Sidi, A.~Nadler, and A.~Shabtai, ``{MaskDGA: A Black-box Evasion Technique
  Against DGA Classifiers and Adversarial Defenses},'' \emph{arXiv preprint
  arXiv:1902.08909}, 2019.

\bibitem{fu2017stealthy}
Y.~Fu, L.~Yu, O.~Hambolu, I.~Ozcelik, B.~Husain, J.~Sun, K.~Sapra, D.~Du, C.~T.
  Beasley, and R.~R. Brooks, ``{Stealthy Domain Generation Algorithms},''
  \emph{IEEE Transactions on Information Forensics and Security}, vol.~12,
  no.~6, pp. 1430--1443, 2017.

\bibitem{pereira2018dictionary}
M.~Pereira, S.~Coleman, B.~Yu, M.~De~Cock, and A.~Nascimento, ``{Dictionary
  Extraction and Detection of Algorithmically Generated Domain Names in Passive
  DNS Traffic},'' in \emph{International Symposium on Research in Attacks,
  Intrusions, and Defenses}.\hskip 1em plus 0.5em minus 0.4em\relax Springer,
  2018, pp. 295--314.

\bibitem{joshi2019using}
A.~Joshi, L.~Lloyd, P.~Westin, and S.~Seethapathy, ``{Using Lexical Features
  for Malicious URL Detection--A Machine Learning Approach},'' \emph{arXiv
  preprint arXiv:1910.06277}, 2019.

\bibitem{yu2019weakly}
B.~Yu, J.~Pan, D.~Gray, J.~Hu, C.~Choudhary, A.~C. Nascimento, and M.~De~Cock,
  ``{Weakly Supervised Deep Learning for the Detection of Domain Generation
  Algorithms},'' \emph{IEEE Access}, vol.~7, pp. 51\,542--51\,556, 2019.

\bibitem{wang2018detection}
Z.~Wang, Z.~Jia, and B.~Zhang, ``{A Detection Scheme for DGA Domain Names Based
  on SVM},'' in \emph{2018 International Conference on Mathematics, Modelling,
  Simulation and Algorithms (MMSA 2018)}.\hskip 1em plus 0.5em minus
  0.4em\relax Atlantis Press, 2018.

\bibitem{kwon2016psybog}
J.~Kwon, J.~Lee, H.~Lee, and A.~Perrig, ``{PsyBoG: A Scalable Botnet Detection
  Method for Large-Scale DNS Traffic},'' \emph{Computer Networks}, vol.~97, pp.
  48--73, 2016.

\bibitem{antonakakis2011detecting}
M.~Antonakakis, R.~Perdisci, W.~Lee, N.~Vasiloglou, and D.~Dagon, ``{Detecting
  Malware Domains at the Upper DNS Hierarchy.}'' in \emph{{USENIX Security
  Symposium}}, vol.~11, 2011, pp. 1--16.

\bibitem{watkins2017using}
L.~Watkins, S.~Beck, J.~Zook, A.~Buczak, J.~Chavis, W.~H. Robinson, J.~A.
  Morales, and S.~Mishra, ``{Using Semi-supervised Machine Learning to Address
  the Big Data Problem in DNS Networks},'' in \emph{2017 IEEE 7th Annual
  Computing and Communication Workshop and Conference (CCWC)}, 2017.

\bibitem{khalil2016discovering}
I.~Khalil, T.~Yu, and B.~Guan, ``{Discovering Malicious Domains through Passive
  DNS Data Graph Analysis},'' in \emph{Proceedings of the 11th ACM on Asia
  Conference on Computer and Communications Security}, 2016, pp. 663--674.

\bibitem{ma2009beyond}
J.~Ma, L.~K. Saul, S.~Savage, and G.~M. Voelker, ``{Beyond Blacklists: Learning
  to Detect Malicious Web Sites from Suspicious URLs},'' in \emph{Proceedings
  of the 15th ACM SIGKDD International Conference on Knowledge Discovery and
  Data Mining}, 2009, pp. 1245--1254.

\bibitem{singh2019detecting}
M.~Singh, M.~Singh, and S.~Kaur, ``{Detecting bot-infected machines using DNS
  fingerprinting},'' \emph{Digital Investigation}, vol.~28, pp. 14--33, 2019.

\bibitem{abbink2017popularity}
J.~Abbink and C.~Doerr, ``{Popularity-based Detection of Domain Generation
  Algorithms},'' in \emph{Proceedings of the 12th International Conference on
  Availability, Reliability and Security}, no.~79.\hskip 1em plus 0.5em minus
  0.4em\relax ACM, 2017.

\end{thebibliography}

\end{document}